\documentclass[final,3p,times]{elsarticle}

\usepackage{amssymb}
\usepackage{amsmath}
\usepackage{soul} 
\usepackage{colortbl}
\definecolor{pink}{rgb}{0.87, 0.36, 0.51}

\usepackage[
    colorlinks=true,
    linkcolor=blue,
    citecolor=blue,
    urlcolor=blue
]{hyperref}

\journal{Journal of Subatomic Particles and Cosmology}

\begin{document}

\begin{frontmatter}



\title{Including Thermal Mesons and Meson Resonances in the Chiral Mean-Field Equation of State}

\author[aaa]{Micheal Kahangirwe}
\author[aaa]{Joaquin Grefa}
\author[ccc]{Mateus Reinke Pelicer}
\author[aaa,ddd]{Rajesh Kumar}
\author[bbb]{Claudia Ratti }
\author[aaa,eee]{Veronica Dexheimer }
\affiliation[aaa]{organization={Center for Nuclear Research, Department of Physics, Kent State University},
             city={Kent},
             postcode={44242},
             state={OH},
             country={USA}}
 \affiliation[ccc]{The Grainger College of Engineering, Illinois Center for Advanced Studies of the Universe,
Department of Physics, University of Illinois at Urbana-Champaign,
             city={Urbana},
             postcode={61801},
             state={IL},
             country={USA}}
 \affiliation[ddd]{Department of Physics, MRPD Government College Talwara,
             city={Punjab},
             postcode={144216},
             country={India}}            
 \affiliation[bbb]{organization={Department of Physics, University of Houston},
             city={Houston},
             postcode={77204},
             state={TX},
             country={USA}}
\affiliation[eee]{organization={CERCA/ISO, Department of Physics, Case Western Reserve University},
             city={Cleveland},
             postcode={44106},
             state={OH},
             country={USA}} 

%% Abstract
\begin{abstract}
The Chiral Mean-Field (CMF) model describes dense matter in terms of baryons and quarks interacting through scalar and vector meson mean fields. In the mean-field approximation, the meson fields are replaced by their expectation values. Their role in generating interactions and in-medium properties is retained, but explicit thermal mesonic excitations are absent. This becomes increasingly problematic at high temperature, where mesons contribute significantly to the thermodynamics of hadronic matter. Here, we keep the original CMF description of dense matter and add the thermal mesonic degrees of freedom using the mesonic sector of the Hadron Resonance Gas (HRG) module in the MUSES framework. Ground-state pseudoscalar and vector mesons account for the missing thermal excitations, while mesonic resonances provide, within the HRG picture, an effective description of mesonic interactions through resonance formation. We study how these contributions affect the pressure, entropy density, energy density, and baryon density, and compare the resulting equation of state with continuum-extrapolated lattice-QCD calculations.
\end{abstract}

%% Keywords

\end{frontmatter}



\section{Introduction}
Developing an equation of state (EoS) that can describe strongly interacting matter across the QCD phase diagram remains a central goal of nuclear physics. Such an EoS should be applicable both to the hot matter produced in relativistic heavy-ion collisions and to the dense matter found in neutron stars and neutron-star mergers at finite temperature \cite{Most:2022wgo,ReinkePelicer:2025vuh,Jahan:2026hvs}. Although these systems can reach comparable baryon densities, they probe very different thermodynamic conditions, particularly in temperature and conserved-charge chemical potentials. A realistic EoS must therefore remain reliable over a broad range of temperatures and densities while satisfying constraints from nuclear experiments, lattice QCD, and astrophysical observations.
The finite-temperature Chiral Mean-Field (CMF) model, based on the nonlinear SU(3) sigma model, provides a description of dense hadronic matter in which baryons interact through scalar and vector meson mean fields~\cite{Dexheimer:2009hi,Cruz-Camacho:2024odu}. The scalar interactions generate effective baryon masses, while the vector interactions modify the effective baryon chemical potentials.
In the mean-field approximation (MFA), only the expectation values of the meson fields are retained. Thermal fluctuations around these mean fields are neglected, so pseudoscalar and vector mesons do not contribute explicitly to the thermodynamic potential. This approximation is well suited to the low-temperature, high-density regime, but becomes less accurate as the temperature increases and thermal mesons become important. As a result, the CMF EoS without explicit thermal mesons underestimates thermodynamic observables at finite temperature compared with continuum-extrapolated lattice-QCD calculations.

In the past, we have also included a free thermal gas of pseudoscalar and vector mesons to CMF~\cite{Dexheimer:2009hi} or, more recently, an interacting gas~\cite{Kumar:2025rxj}.
In this work, we keep the original CMF description of dense matter and add the thermal mesonic degrees of freedom that are missing in the mean-field approximation using the mesonic part of the Hadron Resonance Gas (HRG) module available within MUSES collaboration cyberinfrastructure \cite{Jahan:2026hvs,Vovchenko:2016rkn}. In this way, the dense-matter physics already described by CMF is left unchanged, while the missing thermal meson contribution from HRG naturally incorporates higher-mass mesonic resonances and their finite resonance widths, thereby providing a richer description of the mesonic thermal spectrum. We then study how this extension modifies the thermodynamic observables and compare the results with continuum-extrapolated lattice-QCD calculations.
\begin{figure}[!t]
\centering
\includegraphics[width=0.5\linewidth]{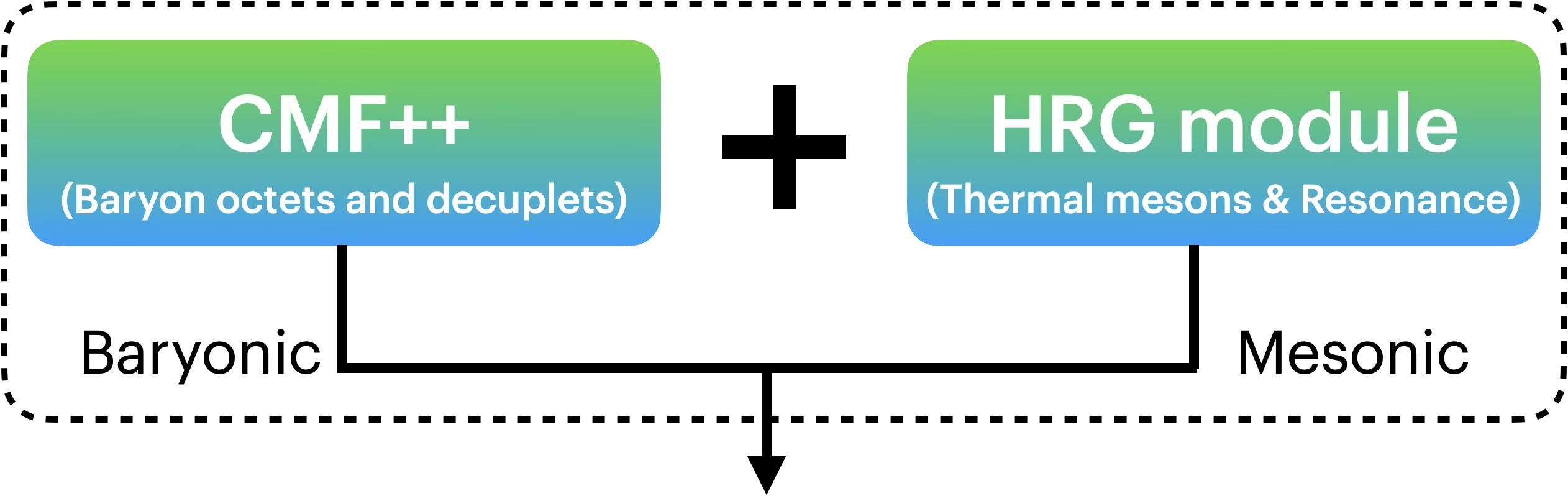}
\caption{Workflow for constructing CMF + thermal mesons EoS within the MUSES framework. The original CMF calculation, including the baryon octet and decuplet, provides the dense-matter contribution, while thermal mesons and mesonic resonances are calculated independently with the HRG module in the ideal-gas limit.}
\label{fig}
\end{figure}

\section{Theoretical Framework}
\label{subsec1}
%\subsection{Chiral Mean-Field Model}

The Chiral Mean-Field (CMF) is based on the nonlinear SU(3) sigma model and provides an effective relativistic description of strongly interacting matter over a broad range of temperatures and densities. Its effective Lagrangian can be written as
\begin{equation}
    \mathcal{L}_{\rm CMF} = \mathcal{L}_{\rm kin} + \mathcal{L}_{\rm int} + \mathcal{L}_{\rm scal} + \mathcal{L}_{\rm vec} + \mathcal{L}_{\rm SB},
\end{equation}
where \(\mathcal{L}_{\rm kin}\) contains the baryonic kinetic terms, \(\mathcal{L}_{\rm int}\) describes baryon–meson interactions, \(\mathcal{L}_{\rm scal}\) and \(\mathcal{L}_{\rm vec}\) contain the scalar- and vector-meson self-interactions, respectively, and \(\mathcal{L}_{\rm SB}\) accounts for explicit chiral-symmetry breaking.
Thermodynamic quantities are obtained in the mean-field approximation for homogeneous, isotropic, and parity-conserving matter. Under these assumptions, the meson fields are replaced by their expectation values: the scalar mean fields and time-like vector components remain nonzero, whereas pseudoscalar fields and spatial vector components vanish. Mesons therefore enter the CMF thermodynamics through their mean-field interactions rather than as explicit thermal excitations. The missing thermal meson contribution is added separately in Sec. \ref{sec:HRG}.

\label{sec:HRG}

\begin{figure}[!t]
    \centering
    \includegraphics[width=0.45\linewidth]{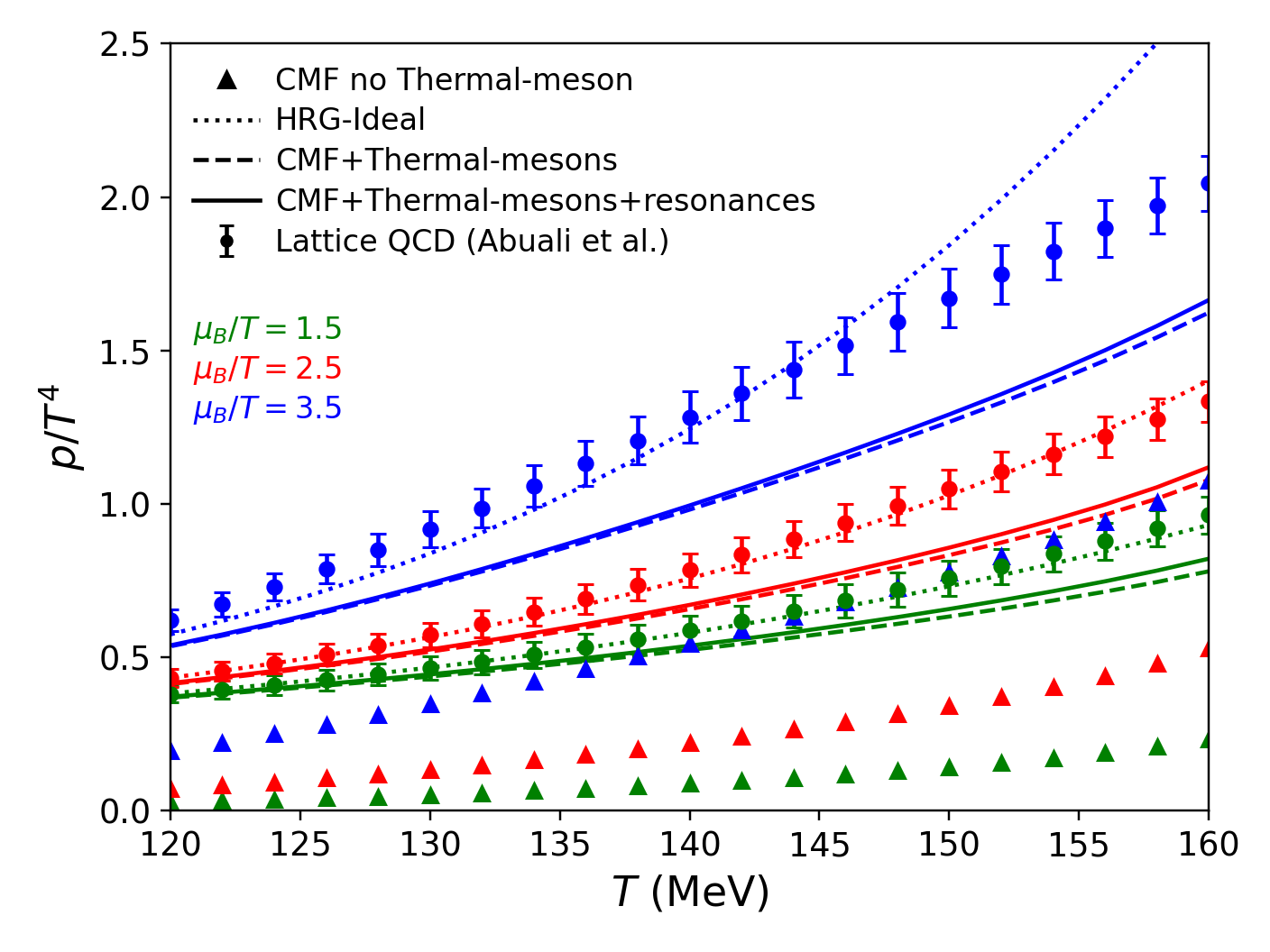}
    \includegraphics[width=0.45\linewidth]{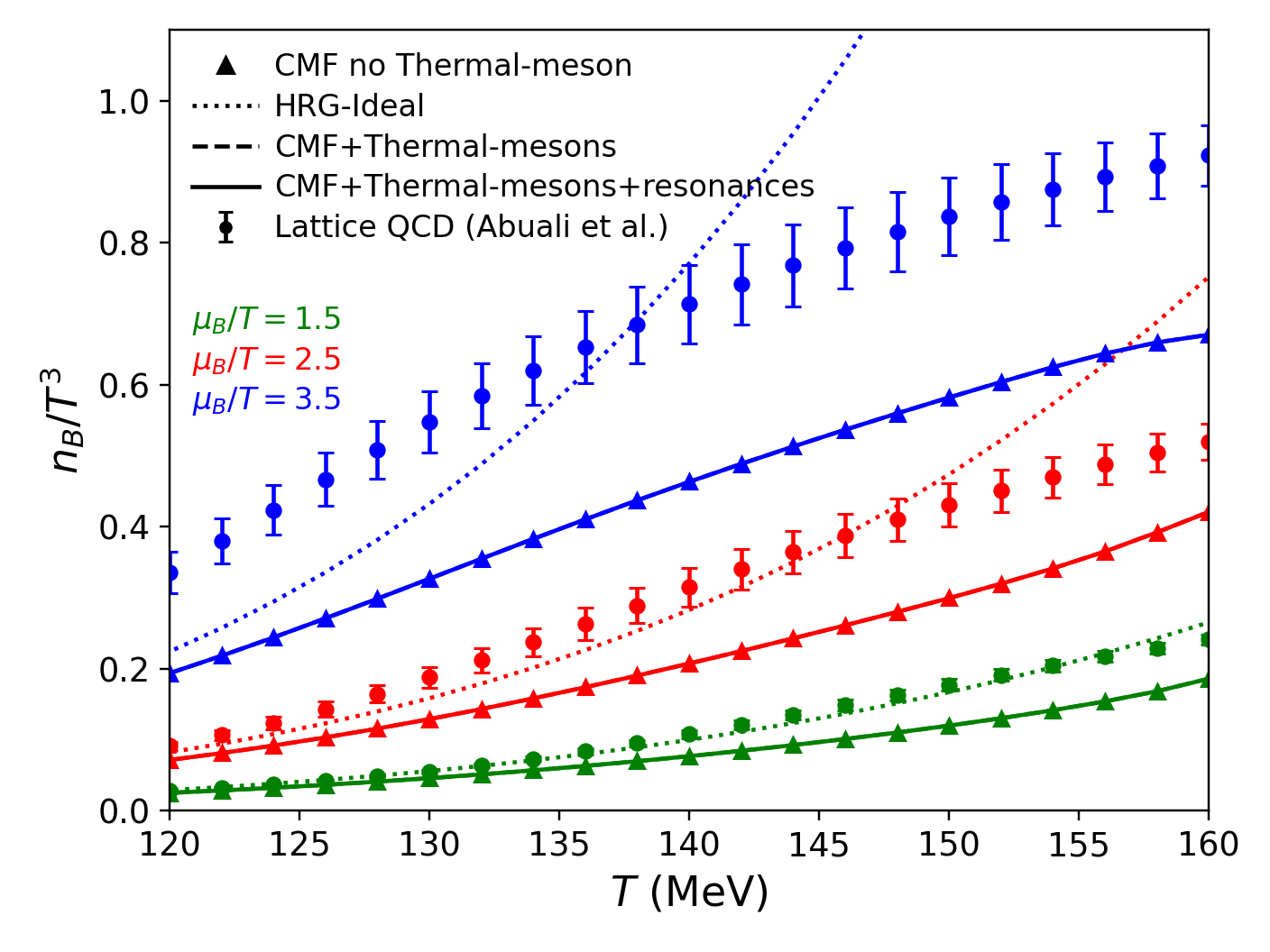}
    \includegraphics[width=0.45\linewidth]{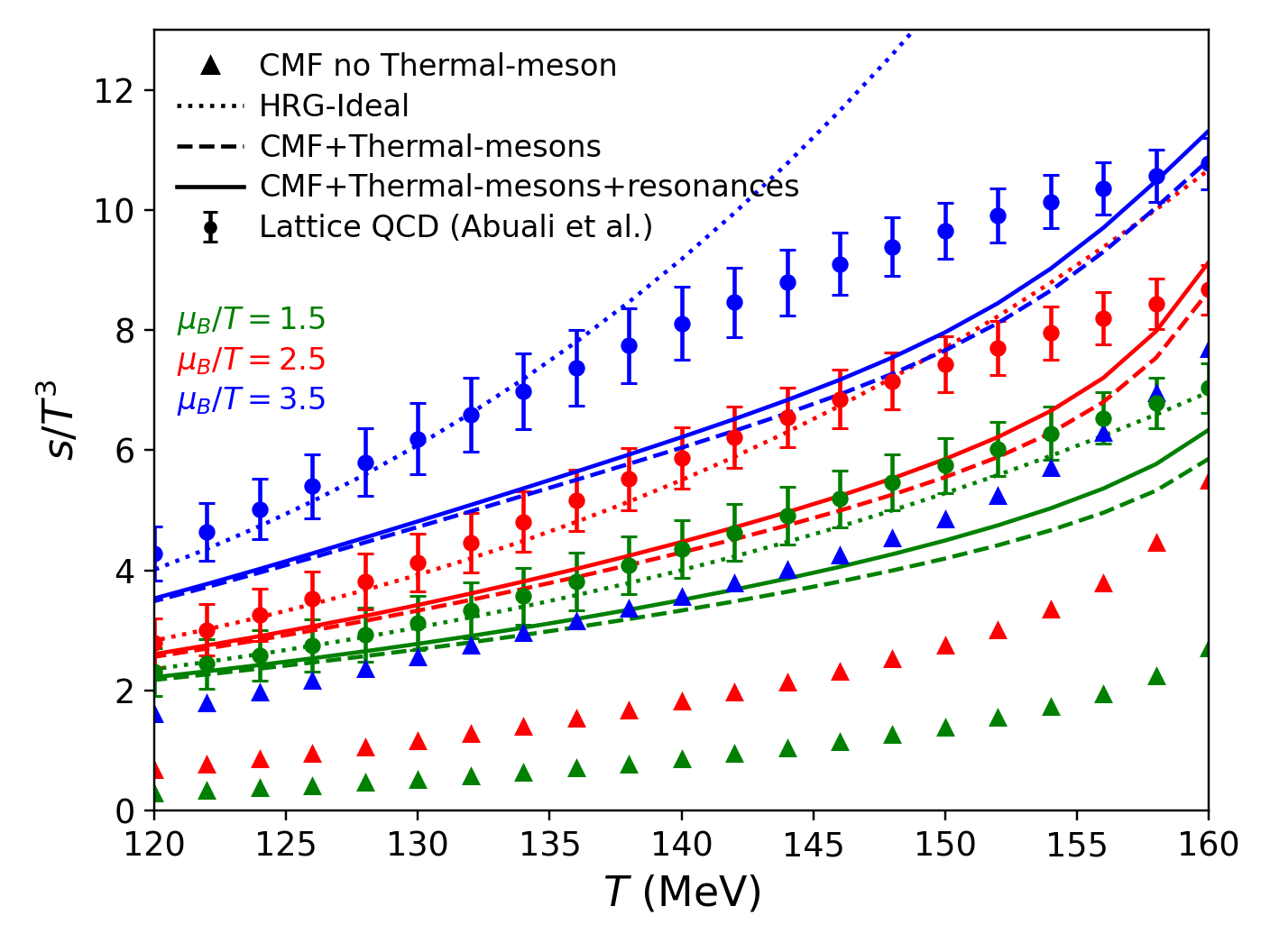}
    \includegraphics[width=0.45\linewidth]{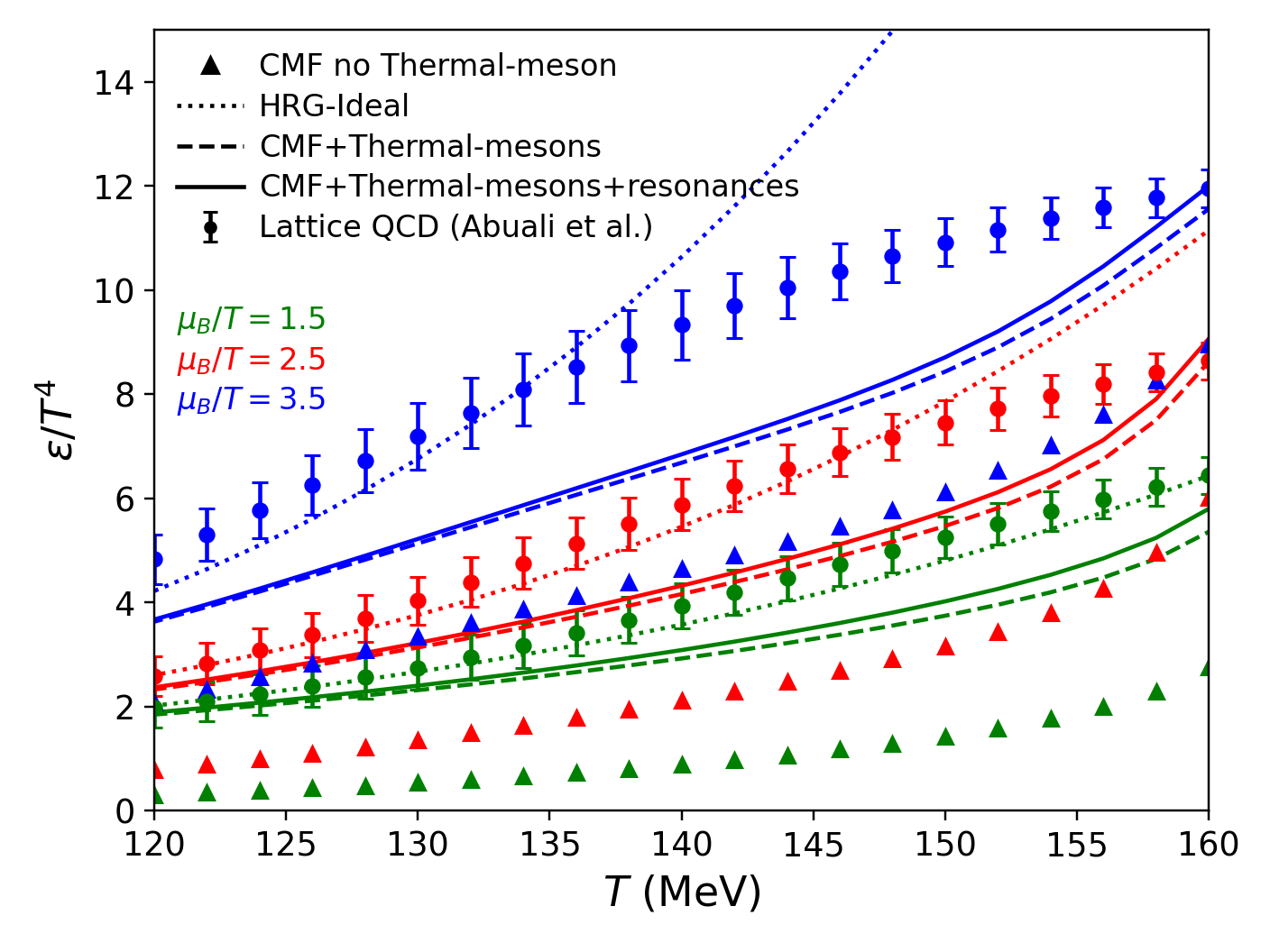}
    \caption{ 
    Thermodynamic observables as functions of temperature for \(\mu_B/T=1.5,\ 2.5,\) and \(3.5\). Triangles show CMF without  thermal mesons, dotted lines show the ideal HRG, dashed lines show CMF with thermal mesons, and solid lines show CMF with thermal mesons and their resonances. Continuum-extrapolated lattice-QCD results \cite{Abuali:2025tbd} are shown as points with error bars. The panels display \(P/T^4\) (top left), \(n_B/T^3\) (top right), \(s/T^3\) (bottom left), and \(\epsilon/T^4\) (bottom right).}
    \label{Thermodynmaics}
\end{figure}

The thermal mesonic contribution missing from the mean-field treatment is calculated using the HRG module in MUSES which is based on Thermal-FIST package \cite{Vovchenko:2019pjl}. We include ground-state pseudoscalar and vector mesons together with their corresponding resonances, all treated as an ideal Bose-Einstein gas. The mesonic pressure is
\begin{equation}
    P_M(T,\mu_i) = \sum_i \frac{g_i T}{2\pi^2} \int_0^\infty k^2\,dk \ln\left[ 1-e^{-(E_i-\mu_i)/T} \right]^{-1},
\end{equation}
where \(g_i\) is the spin-isospin degeneracy,
$$ E_i(k)=\sqrt{k^2+m_i^2}, $$
and \(\mu_i\) is the chemical potential of meson species \(i\).
The ground-state mesons included are
$$ i=\pi,\ K,\ \eta,\ \eta',\ \rho,\ \omega,\ K^*,\ \phi, $$
and the sum is extended to the corresponding pseudoscalar and vector resonances which account for mesonic interactions.

%\subsection{CMF Equation of State with Thermal Meson Contributions}

We consider isospin-symmetric hadronic matter with vanishing electric-charge and strangeness chemical potentials,
$$ \mu_Q=\mu_S=0. $$
The EoS therefore depends only on the temperature \(T\) and baryon chemical potential \(\mu_B\).
The finite-temperature hadronic EoS is constructed by keeping the original CMF contribution and adding the thermal mesonic contribution calculated with the HRG module. The baryon octet and decuplet remain part of the CMF calculation, while the HRG contribution contains only mesons and mesonic resonances, as illustrated in Fig.~\ref{fig}. The total pressure is
\begin{equation}
    P_H(T,\mu_B) = P_{\rm CMF}(T,\mu_B) + P_M(T,\mu_B).
\end{equation}
where, \(P_M\) contains the thermal contribution from the light mesons together with the resonance contribution used to model mesonic interactions effectively.

\section{Results}
Figure~\ref{Thermodynmaics} compares four calculations: CMF without explicit thermal mesons, the ideal HRG model, CMF with ground-state thermal mesons, and CMF with thermal mesons and mesonic resonances. The results are shown for \(\mu_B/T=1.5,\ 2.5,\) and \(3.5\), and are compared with continuum-extrapolated lattice-QCD calculations \cite{Abuali:2025tbd}.
Without explicit thermal mesons, CMF underestimates \(P/T^4\), \(s/T^3\), and \(\epsilon/T^4\), with the difference increasing with temperature. Adding the ground-state pseudoscalar and vector mesons increases these observables and moves the CMF results closer to the lattice-QCD calculations. Including pseudoscalar and vector resonances gives a further contribution and improves the agreement with lattice QCD. Within the HRG, this additional contribution reflects both the larger number of thermally accessible mesonic states and the effective description of mesonic interaction channels through resonance formation.
The baryon density, \(n_B/T^3\), is unchanged when the thermal mesons are added. This is expected because the added mesons carry zero baryon number and, for \(\mu_Q=\mu_S=0\), their pressure does not depend on \(\mu_B\). The baryon density therefore continues to be determined entirely by the original CMF calculation.

\section{Conclusions}
We show that adding explicit thermal mesonic degrees of freedom improves the finite-temperature CMF equation of state without altering the underlying CMF description of dense matter. Ground-state pseudoscalar and vector mesons increase the pressure, entropy density, and energy density, closer to continuum-extrapolated lattice-QCD calculations. In particular, the inclusion of higher-mass mesonic resonances produces a further, although moderate, improvement over the ground-state meson contribution alone, demonstrating that the extended mesonic spectrum provides a relevant correction to the thermodynamics of the hot hadronic phase. As expected, the baryon density is unchanged because the added mesons carry no baryon number. Their contribution also becomes less important at lower temperatures and larger baryon chemical potentials, where the thermodynamics is increasingly controlled by the dense-matter dynamics already captured by the CMF model.
These results represent a step toward extending the CMF equation of state across different thermodynamic regimes of strongly interacting matter. By adding the thermal mesonic contributions that become important at finite temperature, we improve the description of hot hadronic matter relevant to heavy-ion collisions without altering the underlying CMF treatment of dense matter already used for neutron stars and neutron-star mergers. This provides a natural starting point for developing a unified finite-temperature and finite-density EoS that can connect heavy-ion and astrophysical applications. 

\section{Acknowledgments}
This material is based upon work supported by the National Science Foundation under grants No. PHY-2208724, PHY-2116686, PHY-2514763, PHY-2621752 and PHY-2623480, and within the framework of the MUSES collaboration, under Grant No. OAC-2103680. This material is also based upon work supported by the U.S. Department of Energy, Office of Science, Office of Nuclear Physics, under Award Numbers DE-SC0022023 and DE-SC0024700, as well as by the National Aeronautics and Space Agency (NASA) under Award Number 80NSSC24K0767, and the Simons Foundation under award Pivot Fellow-00016159.

% \appendix
% \section{Example Appendix Section}


% %% For citations use: 
% %%       \cite{<label>} ==> [1]

% %%
% Example citation, See \cite{STAR:2005gfr}.

\bibliographystyle{elsarticle-num}
\bibliography{sqm2026_template}



\end{document}